\documentclass[preprint,aps,nofootinbib,tightenlines,
    byrevtex]{revtex4}
\usepackage{graphicx}%
\usepackage{dcolumn}
\usepackage{amsmath}
\usepackage{epsfig}
\usepackage{latexsym}
\usepackage{amssymb}
\usepackage{mathrsfs}

\newcommand{\3}{\ss}

\newcommand{\mpi}{\ensuremath{m_\pi}}

\newcommand{\MeV}{\ensuremath{\mathrm{MeV}}}
\newcommand{\fm}{\ensuremath{\mathrm{fm}}}

\newcommand{\LambdaNoPion}{\ensuremath{\Lambda_{\pi\hskip-0.4em /}}}

 \newcommand{\calD}{\mathcal{D}}
\newcommand{\calH}{\mathcal{H}} \newcommand{\calK}{\mathcal{K}}

\newcommand{\dis}{\displaystyle}
\newcommand{\ii}{\mathrm{i}}
\newcommand{\dd}{\mathrm{d}}

\begin{document}
\bibliographystyle{plain}
\def\nopi{ {\rm EFT}(\pislash) }

\title{Neutron-Deuteron System  and Photon Polarization Parameter at Thermal
Neutron Energies}

\author{H. Sadeghi}\email{H-Sadeghi@araku.ac.ir}
 \affiliation{Department of Physics, University of Arak, P.O.Box 38156-879, Arak,
 Iran.}

\vspace{4cm}

\begin{abstract}
\vspace{0cm}

Effective Field Theory(EFT) is, the unique, model independent and
systematic low-energy version of QCD for processes involving momenta
below the pion mass. A low-energy photo-nuclear observable in
three-body systems,  photon polarization parameter at thermal
neutron energies is calculated by using pionless EFT up to
next-to-next to leading order(N$^2$LO). In order to make a
comparative study of this model, we compared our results for photon
polarization parameter with the realistic Argonne $v_{18}$
two-nucleon and Urbana IX or Tucson-Melbourne three-nucleon
interactions. Three-body currents give small but significant
contributions to some of the observables in the neutron-deuteron
radiative capture cross section at thermal neutron energies. In this
formalism the three-nucleon forces are needed up to N$^2$LO for
cut-off independent results. Our result converges order by order in
low energy expansion and also cut-off independent at this order.

\begin{tabular}{c}
\end{tabular}

\begin{tabular}{rl}

keywords:&\begin{minipage}[t]{11cm} effective field theory,
three-body system, three-body force, photo-nuclear reactions
  \end{minipage}

\end{tabular}
\end{abstract}

\vskip 1.0cm \noindent

\maketitle

\section{Introduction}

The study of the three-body nuclear physics involving
nucleon-deuteron photodisintegration of $^3$He and $^3$H as well as
the time reversed  nucleon radiative capture by deuteron has been
investigated in theoretical and experimental works over the past
decays. The photon polarization in the reaction neutron-deuteron
capture reaction has been measured with polarized thermal neutrons
by Konijnenberg et al.~\cite{Konijnenberg}. In addition the search
for three-nucleon force effects in the electromagnetically induced
process, come more and more into the focus in the recent years. In
principle, using the continuity equation for three-nucleon forces
lead to three-nucleon currents. It is a quantitative question based
on current choices of nuclear force models to reveal signatures by
switching on and off three-nucleon forces.

Several groups are studying electromagnetic processes in the
three-nucleon system. In Refs.~\cite{Golak00,Skib03}, the nucleons
are taken as interacting via two- and three-nucleon potentials and
the electromagnetic currents are then constructed using the exchange
scheme to satisfy the current conservation relation(CCR), but only
with a part of the interaction. In another work, the meson exchange
currents are taken into account using the Siegert's
theorem~\cite{Schad01} without three-body currents.  Three-body
currents are added using a nuclear model which allows for the
excitation of a nucleon to a $\Delta$ isobar. The $\Delta$
excitation yields also effective three-body forces and three-body
currents. However, this current model does not satisfy exactly the
CCR with the adopted Hamiltonian~\cite{Delt04}. In alternative
descriptions of the three-nucleon electromagnetic processes for very
low energy  pd and nd radiative capture, in model dependent theory,
a variety of electromagnetic observables involving the two- and
three-nucleon forces have been extensively studied in the past by
several research groups (for a review, see Ref.~\cite{Car98}).
Recently, Viviani et al. have investigated the nd radiative capture
reactions below deuteron breakup threshold~\cite{Viviani1}. Their
work shows sensitivity to short-range physics namely, details of
including the physics of the Delta and pion-exchange currents. They
obtained the cross section from Argonne $v_{14}$ two-nucleon and
Urbana VIII three-nucleon interactions (AV14/UVIII), also from
Argonne $v_{18}$ two-nucleon and Urbana IX three-nucleon
interactions(AV18/UIX), including $\Delta$ admixtures. They found
Cross section of 0.600 (mb) and 0.578 (mb) which are above the
experimental values by 18$\%$ and 14$\%$, respectively. It is worthy
to mention that the explicit inclusion of $\Delta$-isobar degrees of
freedom in the nuclear wave function improve the agreement with the
experimental data better than  those obtained using the perturbation
theory, $\Delta_{PT}$. This indicate that their results for very-low
energy observables are sensitive to the details of the short-range
part of the interaction. Recent calculations using gauge-invariant
currents reduced the spread~\cite{Marcucci}, however including
three-body currents results $0.558$ mb, which still is above the
data by 10\%. Model-dependent currents associated with
$\Delta(1232)$ were identified as a source of the discrepancy. Thus,
the question that remains is, how such details of short-range
physics can so severely influence a very long range reaction with
maximal energies of less than $10$ MeV.

Recently developed pionless EFT is particularly suited for high
order precision calculation. The so-called pionless EFT in nuclear
physics aspires a systematic classification of all forces. At its
heart lies the tenet that physics at those very low energies can be
described by point-like interactions between nucleons only. In this
approach, all particles but the nucleon themselves are considered
high energy degrees of freedom and are consequently ``integrated
out''. The resulting EFT is considerably simpler than potential
models or the ``pionful'' version of nuclear EFT (in which pions are
kept as explicit degrees of freedom), but its range of validity is
reduced to typical momenta below the pion mass.  There are many
processes situated at thermal energies which are both interesting in
their own right and important for astrophysical applications.
Recently we have calculated the cross section of radiative capture
process $nd\rightarrow {^3H}\gamma $ by using pionless
EFT~\cite{Sadeghi1,Sadeghi2}. No new three-nucleon forces are needed
up to N$^2$LO in order to achieve cut-off independent results, in
addition to those fixed by the triton binding energy and $nd$
scattering length in the triton channel. The cross-section is
determined to be $\sigma_{tot}=[0.503\pm 0.003]mb$.

The present study investigates a low-energy photo-nuclear observable
in three-body systems namely,  photon polarization parameter at
thermal neutron energies, using pionless EFT up to N$^2$LO. The
emphasis is on constructing three-body currents with model
independent theory corresponding to three-nucleon interactions and
comparison of the our model's result with those of other model
dependent theory.

The paper is organized as follows: In Section~\ref{section:nd
radiative capture}, we briefly review theoretical framework
including the Faddeev integral equation, three-body forces and
cut-offs dependence for calculating of neutron-deuteron radiative
capture at thermal energies. Then we calculate the photon
polarization parameter at thermal energies in
Section~\ref{section:Photon polarization parameter} and compare our
results with the corresponding experimental.
Section~\ref{section:comparison} is devoted to comparison of our
result with the data and the results of other theoretical models.
Summary and conclusions are given in
Section~\ref{section:conclusion}.

\section{Neutron-deuteron radiative capture at thermal energies}
\label{section:nd radiative capture}

At thermal energies the $nd$ capture reaction proceeds through
$S$-wave capture predominantly via magnetic dipole transitions from
the initial doublet $J$=1/2 and quartet $J$=3/2 $nd$ scattering
states.  In addition, there is a small contribution due to an
electric quadrupole transition from the initial quartet state.
Consequently, $^2S_{1/2}$ describes the preferred mode for
$nd\rightarrow{^3H}\gamma$ and $pd\rightarrow{^3He}\gamma$.  The
three-nucleon Lagrangian is well-known and will not be discussed
here~\cite{4stooges,griesshammer}.

As long-distance phenomena must be insensitive to details of the
short range physics (and in particular of the regulator chosen),
Bedaque et al.~\cite{4stooges,griesshammer} showed that the system
must be stabilized by a three-body forces
\begin{equation}
  \label{eq:calH}
   \calH(E;\Lambda)=
   \frac{2}{\Lambda^2}\sum\limits_{n=0}^\infty\;H_{2n}(\Lambda)\;
   \left(\frac{ME+\gamma_t^2}{\Lambda^2}\right)^n
   =\frac{2H_0(\Lambda)}{\Lambda^2}+
   \frac{2H_2(\Lambda)}{\Lambda^4}\;(ME+\gamma_t^2)+\dots \;.
\end{equation}
which absorbs all dependence on the cut-off as $\Lambda\to\infty$.
Eq.(\ref{eq:calH}) is analytical in $E$ and can be obtained from a
three-body Lagrangian, employing a three-nucleon auxiliary field
analogous to the treatment of the two-nucleon
channels~\cite{4stooges}. Contrary to the terms without derivatives,
the term involves three-body forces (second term) contains two
derivatives. The derivation of the integral equation describing
neutron-deuteron scattering has been discussed
before~\cite{griesshammer}. We present here only the result,
including the new term generated by the second term in
Eq.(\ref{eq:calH}).  The resulting amplitudes is a mixture of $t_s$
describes the $d_t + N\rightarrow d_s + N$ process, and $t_t$
describes the $d_t + N\rightarrow d_t + N$ process:

\begin{eqnarray}
\label{eq:Int}
 t_s(p,k)& =  \frac{1}{4}\left[3\mathcal{K}(p,k)
+2\mathcal{H}(E,\Lambda)\right]+\dis\frac{1}{2\pi}
 \int\limits_0^\Lambda \dd q\; q^2&\left[\mathcal{D}_s(q)\left[\mathcal{K}(p,q)+2\mathcal{H}(E,\Lambda)
      \right]
t_s(q)\right.\nonumber\\
       &\left.+\mathcal{D}_t(q)\left[3\mathcal{K}(p,q)+2\mathcal{H}(E,\Lambda)
       \right]
t_t(q)\right] \label{int_equation_triton}\nonumber\\
 t_t(p,k)& = \frac{1}{4}\left[\mathcal{K}(p,k)
+2\mathcal{H}(E,\Lambda)\right]+\dis\frac{1}{2\pi}
 \int\limits_0^\Lambda \dd q\; q^2&\left[ \mathcal{D}_t(q)\left[
\mathcal{K}(p,q)+2\mathcal{H}(E,\Lambda)\right]
t_t(q)\right.\nonumber\\
       & & \left.+\mathcal{D}_s(q)
       \left[3\mathcal{K}(p,q)+2\mathcal{H}(E,\Lambda)\right]
t_s(q)\right]\;\;,
\end{eqnarray}
where $\mathcal{D}_{s,t}(q)=\mathcal{D}_{s,t}(E-\frac{q^2}{2M},q)$
are the propagators of deuteron.  For the spin-triplet
$\mathrm{S}$-wave channel, one replaces the two boson binding
momentum $\gamma$ and effective range $\rho$ by the deuteron binding
momentum $\gamma_t=45.7025\;\mathrm{MeV}$ and effective range
$\rho_t=1.764\;\mathrm{fm}$. Because there is no real bound state in
the spin singlet channel of the two-nucleon system, it is better to
determine the free parameters by the scattering length
$a_s=1/\gamma_s=-23.714\;\mathrm{fm}$ and the effective range
$r_s=2.73\;\mathrm{fm}$ at zero
momentum~\cite{4stooges,griesshammer}. The neutron-deuteron $J=1/2$
phase shifts $\delta$ is determined by the on-shell amplitude
$t_t(k,k)$, multiplied by the wave function renormalisation
\begin{equation}
T(k)=Z t_t(k,k)=\frac{3\pi}{M}\frac{1}{k \cot\delta-\ii k}\;\;.
\end{equation}

The spine structure of the matrix elements  for neutron radiative
capture by deuteron is complicated, however in very low energy for
this reaction we can introduced three multipole transition that is
allowed by p-parity and angular momentum conservation i.e.
$I^p={\frac{1}{2}}^+\rightarrow M_1$ and
$I^p={\frac{3}{2}}^+\rightarrow M_1, E_2$. The parameterization of
the corresponding contribution to the matrix elements and the
$\mathcal{M}_1$ amplitude are from the magnetic moments of the
nucleon and dibriyon. These are well-known and will not be given
here~\cite{Sadeghi1,Sadeghi2}.

The radiative capture cross section $nd\rightarrow ^3H\gamma$ at
very low energy is given by~\cite{Sadeghi1},

\begin{equation}\label{crosssection1}
  \sigma=\frac{2}{9}\frac{\alpha}{v_{rel}}\frac{p^3}{4M^2_N}\sum_{iLSJ}
  [{|\widetilde{\chi}^{LSJ}_i|}^2]
\end{equation}
where
\begin{equation}\label{redefine}
  \widetilde{\chi}^{LSJ}_i=\frac{\sqrt{6\pi}}{p\mu_N} \sqrt{4\pi} {\chi^{LSJ}_i}
\end{equation}
where $\chi$ is either the magnetic or electrical moment and $\mu_N$
is nuclear magneton and p is momentum of the incident neutron in the
center of mass. The contribution of the electric transition
$E^{LSJ}_i$ to total cross section at energies less than 60 KeV is
insignificant. Therefore, the electric quadrupole transition $E^{0
(3/2) (3/2)}_2$ from the initial quartet state will not be
considered at thermal energies.

\begin{figure}[!t]
\begin{center}
 \includegraphics*[width=.5\textwidth]{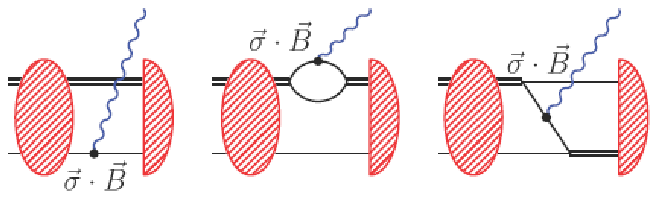}
 \caption{(Color online) Some diagrams for adding photon-interaction to the Faddeev equation
   up to N$^2$LO. Thick solid
  line is propagator of the two intermediate auxiliary fields $D_s$ and $D_t$,
  denoted by $\calD$; $\calK$:
  propagator of the exchanged nucleon. Photon is minimally coupled to nucleons in three-body systems.
   Wavy line shows photon and small circles show magnetic
   photon interaction.} \label{fig1}
\end{center}
\end{figure}

\begin{figure}[!t]
\begin{center}
 \includegraphics*[width=.6\textwidth]{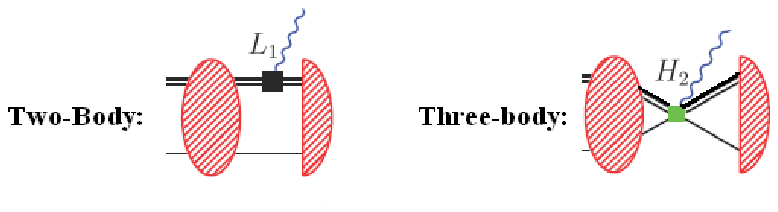}
 \caption{(Color online) Some diagrams for adding photon-interaction to the Faddeev equation
   up to N$^2$LO.  Photon is minimally coupled to two-body system and three-body vertices in three-body systems.
   For $L_1$ vertices, see Ref~\cite{Rupak98}; $H_2$:three- body
   force, see eq.(1).  Remaining notation as in Fig.~\ref{fig1}.} \label{fig2}
\end{center}
\end{figure}

We now turn to the Faddeev integral equation used in the magnetic
moment calculation and also the interaction kernel included in this
integral equation. Fig.~\ref{fig1} represents the contribution
diagrams for adding photon-interaction to the Faddeev equation
(\ref{eq:Int}). In these diagrams photon is minimally coupled to
nucleons in three-body systems. The diagrams for adding
photon-interaction to the Faddeev equation up to N$^2$LO are
depicted in Fig.~\ref{fig2}. Photon is coupled to two-body system
via $L_1$ vertices. The coefficient $L_1$ is fixed at its leading
non-vanishing order by the thermal cross section~\cite{Rupak98}. We
have other possible diagram that can be considered for our
calculation for inclusion of photon to the three-body vertices
$\mathcal{H}$. This diagram is shown in Fig.~\ref{fig2}.

All corrections contribute to observables typically as
  $  Q^{n}=
  \left(\frac{p_\mathrm{typ}}{\LambdaNoPion}\right)^{n}$
compared to the LO result and that low-energy observables must be
independent of an arbitrary regulator $\Lambda$ up to the order of
the expansion. In other words, the physical scattering amplitude
must be dominated by integrations over off-shell momenta $q$ in the
region in which the EFT is applicable, $q\lesssim\LambdaNoPion$.
Typical low momentum scales $p_\mathrm{typ}$ in the three-body
system are the binding momenta of the two-nucleon real and virtual
bound states, $\gamma_s\approx -8.0\;\MeV,\;\gamma_t\approx
45\;\MeV$ and the scattering momentum $k$. In addition, the
three-body forces are determined in part by the typical
three-nucleon bound state momentum $\gamma_d\sim\sqrt{MB_d}\approx
90\;\MeV$, $B_d$ is the triton binding energy. The breakdown scale
$\LambdaNoPion\approx\mpi$ of the theory is the scale at which
higher order corrections become comparable in size. One can
therefore estimate sensitivity to short-distance physics, and hence
provide a reasonable error analysis, by employing a momentum cut-off
$\Lambda$ in the solution of the Faddeev equation and varying it
between the breakdown-scale $\LambdaNoPion$ to $\infty$. If
observables change over this range by ``considerably'' more than
$Q^{n+1}$, a counter-term of order $Q^n$ should be added. This
method is frequently used to check the power counting and systematic
errors in pionless EFT with three nucleons, see e.g.~most
recently~\cite{griesshammer}. A similar argument was also developed
in the context of the EFT ``with pions'' in nuclear
physics~\cite{Bernard:2003rp,Epelbaum:2003gr}.

\begin{table}[!t]
\caption{Comparison between different theoretical and experimental
results for Neutron radiative capture by deuteron at zero energy
(0.0253 ev). Last row shows N$^2$LO order pionless EFT result. }
\label{tab:a} \vspace{0.25cm}
\begin{center}
\begin{tabular}{c c||c|c}
\hline & Experiment & Year & Total cross section(mb)  \\
 \hline \hline
 & Jurney et.al.~\cite{Jurney63} & 1963 & 0.60 $\pm$ 0.05\\
 & Merritt et.al.~\cite{Merritt} & 1968 & 0.521 $\pm$ 0.009\\
 & Jurney et.al.~\cite{Jurney} & 1982 & 0.508 $\pm$ 0.015 \\
\hline
 & Theory &  &   \\
 \hline \hline
 & AV14/VIII(IA+MI+MD+$\Delta$) ~\cite{Viviani1}& 1996 & 0.600  \\
 & AV18/IX(IA+MI+MD+$\Delta$)~\cite{Viviani1} & 1996 & 0.578   \\
 & AV18/IX (gauge inv.)~\cite{Marcucci} & 2005 & 0.523\\
 & AV18/IX (gauge inv. + 3N-current)~\cite{Marcucci} & 2005 & 0.556\\
 & EFT(N$^2$LO)+3N-forces~\cite{Sadeghi2} & 2006 &$0.503\pm0.003$     \\

\hline
\end{tabular}
\end{center}
\end{table}

\begin{figure}[!htb]
\begin{center}
  \includegraphics[width=0.7\linewidth,clip=true]{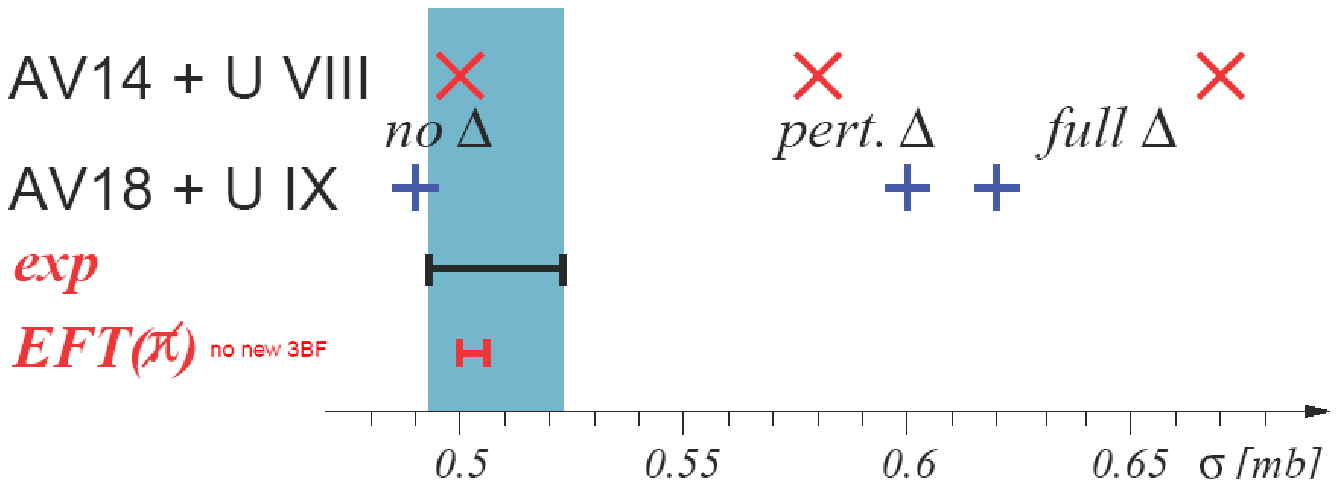}
\end{center}
\vspace*{-0pt} \caption{(Color online) Comparison between calculated
cross section of neutron radiative capture by deuteron by different
theoretical models, pionless EFT and experimental data.}
\label{crosssection}
\end{figure}

\section{Photon polarization parameter at thermal energies}
\label{section:Photon polarization parameter}
If the process is dominated by S-wave capture, as in the case for
the neutron-deuteron radiative capture reaction at thermal neutron
energies, the observable for circular polarization
$P_\Gamma(\theta)$ is simply given by:
\begin{equation}\label{Photon polarization}
  P_\Gamma(\theta)=R_c {P_N} \cdot {\hat{q}}
\end{equation}
where $P_N$ is the polarizations of the spin-1/2 nucleon  and $R_c$
is the polarization parameter (for more detail see~\cite{Viviani1}).
This polarization parameter depends on the relative sign between the
amplitudes 1/2 and 3/2 channels. Numerically, $R_c$ lies in the
region $-1/2 \leq R_c \leq 1$. Its experimental value is $R_c$ =
-0.42 ~\cite{Konijnenberg}.

\section{Results}
\label{section:comparison}

We numerically solved the Faddeev integral equation up to N$^2$LO.
We used $\hbar c=197.327\;\MeV\,\fm$, and  mass of $M=938.918\;\MeV$
for nucleon. A deuteron binding energy (momentum) of $B=2.225\;\MeV$
($\gamma_d=45.7066\;\MeV$) is used for the nucleon-nucleon triplet
channel. A residue of $Z_d=1.690(3)$ is used for the $NN$ singlet
channel. The  ${}^1\mathrm{S}_0$ scattering length is chosen to be
$a_s=-23.714\;\fm$. After fixing the leading non-vanishing order in
the thermal cross section $L_1$ is found to be -4.5 fm.

As in Ref.~\cite{Sadeghi2}, we determined which three-body forces
are required at any given order, and how they depend on the cut-off.
Low energy observables must be insensitive to the cut-off, namely to
any details of short range physics in the region above the break
down scale of the pionless EFT ( which set approximately by the pion
mass ).

The results for the thermal energy cross section and photon
polarization parameter are presented in table {I} and {II}, along
with the experimental data~\cite{Jurney,Konijnenberg}. Table {I}
compares the $nd\rightarrow ^3H\gamma$ cross section at zero energy
(0.0253 eV) for various experimental and theoretical works. The
corresponding values for the cross section from the pionless EFT
evaluation up to N$2$LO is shown in the last row. The EFT results
for this cross section are presented  only up to  three significant
digits.

Recently in a model dependent two-body current calculation, the
total cross section for $nd\rightarrow ^3H\gamma$ is obtained as
$\sigma_T=0.523$ mb~\cite{Marcucci}. This value can be compared with
the corresponding result $\sigma_T=0.558$ mb obtained in
Ref.~\cite{Viviani1}. The later work used the present MI two-body
current operators therefore leads to an estimate closer to the
experimental data $\sigma_T=0.508\pm015$ mb~\cite{Jurney}. However,
the addition of the three-body currents, which give a rather sizable
contribution as can be seen from the row labeled ``full-new'' in
table {I}, brings the total cross section to $\sigma_T=0.556$ mb.
Table {I} shows also EFT result of the cross section for this
reaction up to N$^2$LO order. Three-nucleon forces are needed up to
N$^2$LO order for cut-off independent results. Hence the
cross-section is in total determined as
$\sigma_{tot}=[0.485(LO)+0.011(NLO)+0.007(N^2LO)]=[0.503\pm
0.003]mb$. The theoretical accuracy may for example be estimated
conservatively by $Q\sim\frac{\gamma_t}{\mpi}\approx\frac{1}{3}$ of
the difference between the NLO and N$^2$LO results.

Table {II}  shows Comparison between the results of different models
dependent, model independent EFT and experiment for the photon
polarization parameter. The photon polarization parameter is
sensitive to two-body currents ( for its definition in terms of
RMEs, see Ref.~\cite{Viviani1}).  We compare our prediction for the
photon polarization parameter with the theoretical and the
experimental results of Ref.~\cite{Viviani1,Marcucci,Konijnenberg}
in this table. The magnetic $M_1$-transition gives the dominant
contribution for our calculation.

In Fig.~\ref{crosssection}, we compare our results with those
obtained in Refs.~\cite{Viviani1,Jurney}. As can be seen by
inspecting Fig.~\ref{crosssection}, the pionless EFT calculations is
converges order by order in low energy expansion and also cut-off
independent at this order(see~\cite{Sadeghi2}). There are no new
three-nucleon forces besides those already fixed in $nd$ scattering
at the same order. The contribution from the photon coupling to a
three-nucleon force is negligible in our calculation. Our
calculation has a systematic error which is now smaller than the
experimental error-bar.

\begin{table}[!htb]
\caption{Comparison between different theoretical results for photon
polarization parameter $R_C$ of the reaction neutron-deuteron
radiative capture at thermal energies. The last line quotes
deviation between data and theory, if it is larger than the
theoretical or experimental uncertainty.} \label{tab1}
\vspace{0.25cm}
\begin{center}
  \begin{tabular}{c||c|c c}
   \hline Type of theory and experiment  & $R_C$ & Overestimation the experimental  \\
    \hline \hline
    AV14/VIII(IA+MI+MD+$\Delta$) ~\cite{Viviani1}      & -0.420 &  below 1$\%$    \\
    AV18/IX(IA+MI+MD+$\Delta$)~\cite{Viviani1}         & -0.469 &  12$\%$    \\
    AV18/IX (gauge inv.+ 2N-current)~\cite{Marcucci}  & -0.429 &  2$\%$    \\
    AV18/IX (gauge inv. + 3N-current) ~\cite{Marcucci} & -0.476 &  13$\%$   \\
    EFT(LO)                                           & -0.387 &  8$\%$   \\
    EFT(NLO)                                          & -0.403 &   4$\%$   \\
    EFT(N$^2$LO)                                      & -0.412 &   2$\%$   \\
\hline
    Experiment~\cite{Konijnenberg}  & $-0.42\pm0.03$ \\
    \hline
\end{tabular}
\end{center}
\end{table}

\section{Summery and Conclusion}
\label{section:conclusion}
We applied pionless EFT to find numerical results for the photon
polarization parameter $R_C$. At very low energies, the interactions
between nucleons can be described only by point-like interactions.
One cannot identify pions as the lightest exchanged particles
between nucleons as long as the typical external momentum
$p_\mathrm{typ}$ in a reaction is below the pion mass $\mpi$. That
is because the Compton wavelengths are not small enough to resolve
the nuclear forces as originating in part from one pion exchange.
Then all particles but nucleons are integrated out. One can identify
a small, dimensionless parameter
$Q=\frac{p_\mathrm{typ}}{\LambdaNoPion}\ll 1$, where
$\LambdaNoPion\sim\mpi$ is the typical momentum scale at which the
one pion exchange is resolved and pionless EFT must break down.
Incident thermal neutron energies have been considered for this
capture process.

The photon polarization parameter $R_c$ of the reaction
neutron-deuteron radiative capture $nd\rightarrow \gamma{^3H}$ at
thermal energies was calculated in pionless EFT. This model
independent and systematic low energy version of QCD is suited for
processes involving momenta below the pion mass. At these energies
our calculation is dominated by  $S$-wave state and magnetic
transition $M_1$ contribution only. The $M_1$ amplitude is
calculated up to N$^2$LO. Three-nucleon forces are needed up to
N$^2$LO order for cut-off independent results. The triton binding
energy and nd scattering length in the triton channel have been used
to fix them. Hence the The photon polarization parameter in total is
determined as
$R_c=-[0.387(LO)+0.016(NLO)+0.009(N^2LO)]=[-0.412\pm0.003]$. This
converges order by order in low energy expansion and also is cut-off
independent at this order. We notice that our calculation has a
systematic error which is now smaller than the experimental error
bar.

\section{Acknowledgments}
I would like to thank Harald W. Grie\ss hammer for useful and
valuable comments and L. Marcucci for providing the recent
theoretical results. This work was performed  under the auspices of
the university of Arak under contract No.~11-18680.


\end{document}